\documentclass[a4paper,11pt]{article}
\usepackage{jinstpub}
\usepackage{subfigure}
\usepackage{xfrac}
\usepackage{lineno}

\title{Simulations of the nEDM@SNS light collection system efficiency}

\author[a,1,2]{D. A. Loomis,\note{Corresponding author.}\note{Current Address: University of Michigan, Ann Arbor, MI 48104, USA.}}
\author[b]{V. Cianciolo,}
\author[c]{E. Leggett}

\affiliation[a]{Western Kentucky University,\\ Bowling Green, KY 42101, USA}
\affiliation[b]{Physics Division, Oak Ridge National Laboratory,\\ Oak Ridge, TN 37831, USA}
\affiliation[c]{Mississippi State University,\\ Starkville, MS 39762, USA}

\emailAdd{dloom@umich.edu}

\abstract{A system for collecting the scintillation light produced by the capture process of ultra-cold neutrons (UCN) on polarized \textsuperscript{3}He is discussed and results from simulations of its performance are presented. This system will be implemented in nEDM@SNS, the experiment searching for the neutron electric dipole moment (nEDM) at the Spallation Neutron Source (SNS) at Oak Ridge National Laboratory. Simulation results show that the light collection system collects on average 17 photoelectrons per UCN-\textsuperscript{3}He capture event (sufficient to generate a robust signal), reconstructs the event location in the beam direction to approximately 3 cm accuracy, detects capture events with a high and spatially uniform efficiency (0.95 with 1\% variation), and rejects greater than 50\% of beta decay background events.}

\keywords{Simulation methods and programs, Photon detectors for UV, visible and IR photons (vacuum) (photomultipliers, HPDs, others), Scintillators, scintillation and light emission processes (solid, gas and liquid scintillators)}

\begin{document}
\maketitle
\flushbottom

\section{Introduction} \label{sec:introduction}
The presence of a nonzero permanent neutron electric dipole moment (nEDM), a measure of the charge distribution within the neutron, would be a violation of both parity (P) and time reversal (T) invariance and a clear signature of physics beyond the standard model.
The nEDM@SNS experiment \cite{Ahmed} will utilize the technique introduced by Golub and Lamoreaux \cite{Golub} to measure the nEDM with a two order of magnitude improvement of sensitivity over recent nEDM measurements \cite{PSI}. Neutrons with de Broglie wavelength $\lambda = 8.9 \textrm{\AA}$ (kinetic energy = 1 meV) from the Fundamental Neutron Physics Beamline at the ORNL Spallation Neutron Source are directed into two measurement cells filled with superfluid \textsuperscript{4}He and a dilute concentration of polarized \textsuperscript{3}He.   
The superfluid \textsuperscript{4}He downscatters a fraction of the incoming neutrons to energies low enough that they can be trapped within the measurement cells. These "ultracold neutrons" (UCN) can be stored in the measurement cells for long periods (hundreds of seconds \cite{Leung}), maximizing observation time and increasing sensitivity. Both neutrons and \textsuperscript{3}He have a magnetic dipole moment and will precess in a magnetic field.
The neutron-\textsuperscript{3}He capture process
\begin{equation} \label{eq:1}
n + {}^{3}He \xrightarrow{} p + {}^{3}H + 764 \: \textrm{keV}
 \end{equation}
deposits 764 keV of kinetic energy into the superfluid \textsuperscript{4}He, which in turn produces a burst of extreme ultraviolet (EUV) scintillation light (80 nm) \cite{Fleishman, Thorndike, McKinsey}. 
The capture process is strongly spin-dependent. As a result, the scintillation rate will vary as $1-P_3 P_n (\hat{n} \cdot \hat{3})$, where $P_3,P_n$ and $\hat{n},\hat{3}$ are the polarization and spin direction of the two species. The nEDM can be extracted with two different measurement techniques ("free-precession" and "spin dressing"), both of which use the scintillation signal resulting from neutron/\textsuperscript{3}He capture events to monitor the angle between neutron and \textsuperscript{3}He spin vectors.

To measure the scintillation signal from UCN-\textsuperscript{3}He capture, a light collection system, composed of a wavelength-shifting (WLS) coating and thin dielectric reflector embedded in the measurement cell walls, WLS optical fibers, and an array of silicon photomultipliers (SiPM’s) (a more thorough description of the system is offered in section \ref{sec:methods}), has been designed in accordance with the experiment requirements. Many design choices were made prior to this study. For instance, due to the need for flexibility of the system in the face of thermal contraction and overall routing complexity, it was chosen to be fiber-based. Given the emission spectrum of the WLS coating, Kuraray Y-11 fibers were selected. Electrostatic requirements constrain the fiber placement to be adjacent to the ground electrode of the high voltage system. SiPM's were chosen due to their high photoelectron detection efficiency (PDE) and sufficient dark rate suppression is obtained by cooling them.

The simulation study described in this report was undertaken to address remaining design parameters whose impact can be evaluated with a Monte Carlo approach. These parameters include dielectric reflector surface roughness, fiber dye density, diameter, and length, and the choice of embedding the fibers in the measurement cell walls. With these parameters addressed, we then carried out a capture event simulation to quantify the light collection system’s event detection and beta decay background rejection efficiencies, as well as event position reconstruction capabilities. 

In section \ref{sec:methods} we discuss the light collection apparatus in more detail, and the simulation framework used in this study. In section \ref{sec:results}, we discuss simulation results, in particular showing the efficiency of signal detection and beta decay background rejection. Finally, in section \ref{sec:conclusions} we present a summary of the findings.

\section{Apparatus and Procedure} \label{sec:methods}
\subsection{Light collection apparatus} \label{sec:lightcollectionapp}
\begin{figure*}[htbp!]
    \centering
    \subfigure[]{\label{fig:CentralVolume}\includegraphics[width=0.49\textwidth]{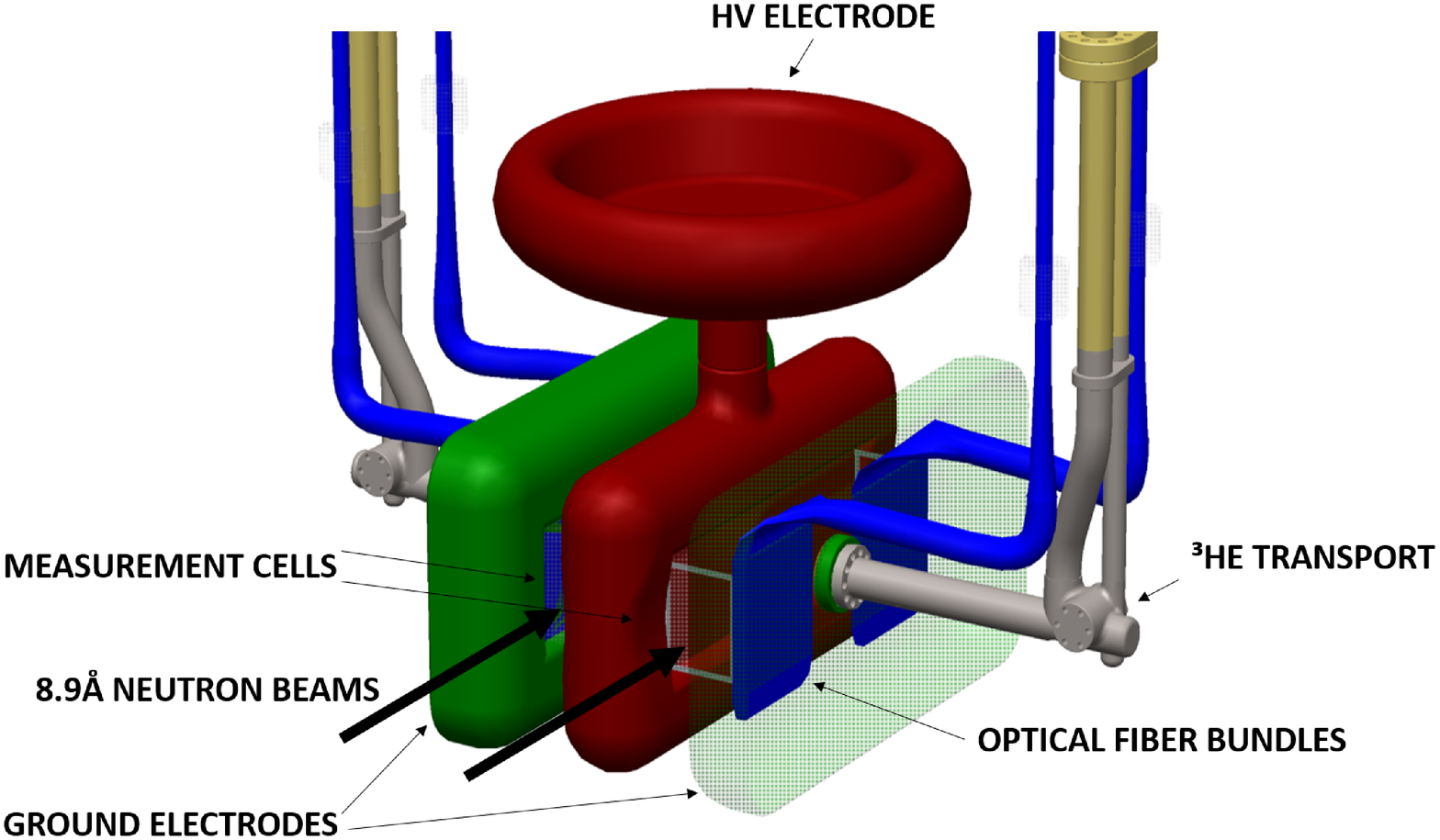}}
    \subfigure[]{\label{fig:LightCollectionSystem}\includegraphics[width=0.49\textwidth]{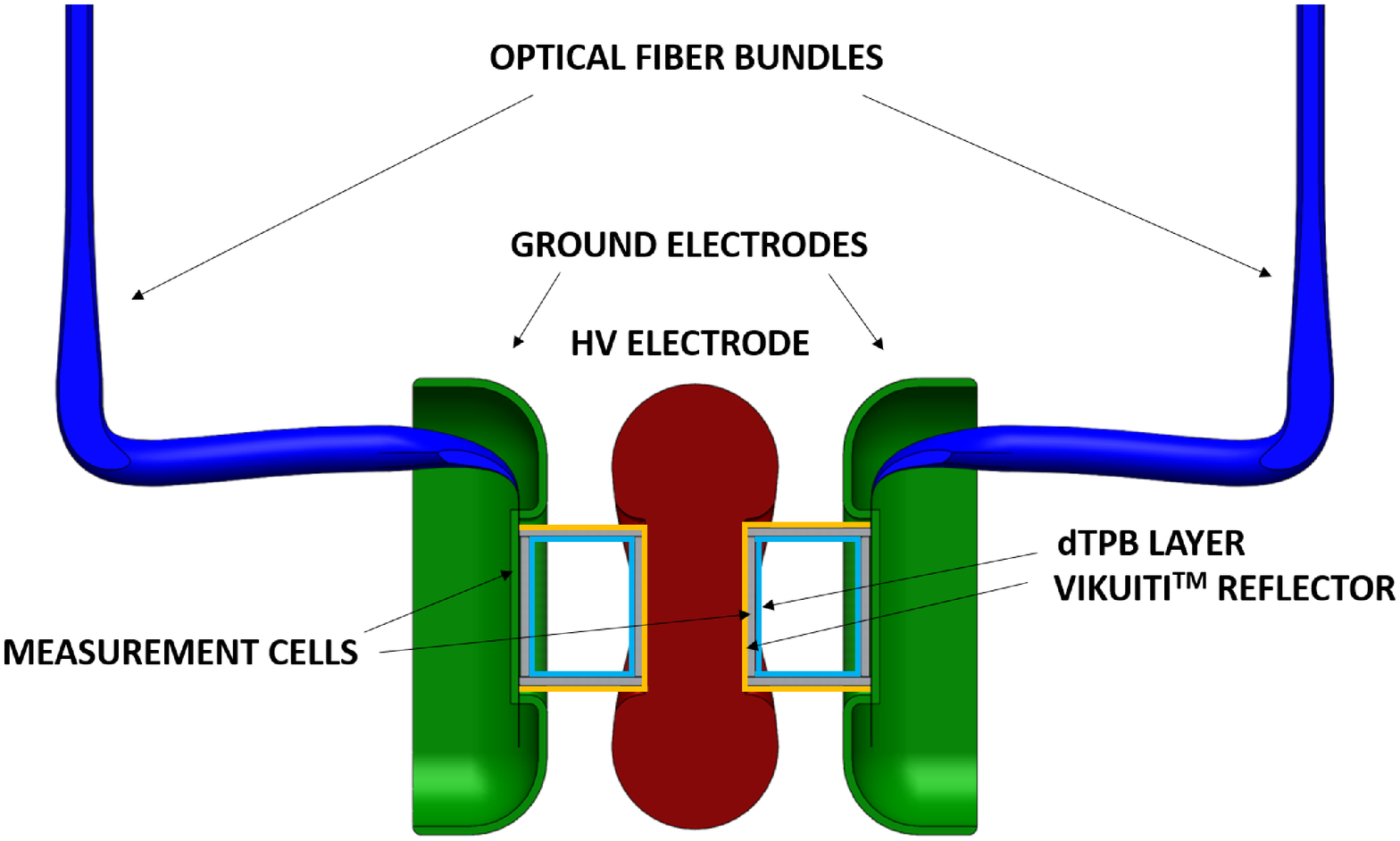}}
    \caption{(a) Components inside the nEDM@SNS Central Volume \cite{Ahmed}. (b) Beam's-eye view highlighting components of the light collection system.}
    \label{fig:Apparatus}
\end{figure*}

The light collection system, located inside the experiment’s central volume with the measurement cells and high voltage electrodes (figure \ref{fig:Apparatus}), is responsible for transporting the scintillation light from the UCN-\textsuperscript{3}He capture events to the readout electronics located exterior to the volume.

EUV photons have short absorption lengths in nearly all materials other than helium, so they must be converted into optical wavelengths by a WLS coating on the measurement cell walls. A coating made by dissolving deuterated tetraphenyl butadiene (dTPB) \cite{McKinsey2,Habicht,Gehman} and deuterated polystyrene (dPS) \cite{Lamoreaux} in deuterated toluene has been shown to retain the \textsuperscript{3}He polarization \cite{Ye} and neutron density \cite{Leung} for times that are long relative to the neutron beta decay lifetime. Some of the photons that enter the dTPB coating as EUV photons are converted to blue wavelengths and collected by Kuraray Y-11 WLS fibers pressed against the measurement cell wall near the ground electrode. To prevent a substantial number of photons from escaping the measurement cell before reaching the optical fibers, a dielectric reflector film (Vikuiti VM2000) will be glued to the other cell walls. 

Once in the fibers, the photons are converted to green wavelengths and transported to the surface of the central volume. There they are mated to clear fibers which lead to an array of silicon photomultipliers (SiPMs) that are placed outside the cryovessel due to the strict non-magnetic requirements near the central volume. Hamamatsu 13360-75 SiPMs \cite{Hamamatsu} were selected for their high PDE, their small area well-matched to individual Y-11 fibers (dark rate is proportional to area), and because they are well characterized at the low temperatures needed (< -40 C) to further suppress dark rate \cite{Nepomuk}. 

A photon counting technique \cite{Ahmed}, in which any SiPM detection is assumed to correspond to a single photon, is implemented by discriminating each analog SiPM readout at $\sim0.5$ per photoelectron to create a digital signal. With this technique, we record the arrival time of each photon, enabling offline pulse-shape discrimination. Additionally, this technique eliminates optical cross talk, allowing the SiPMs to be operated at a relatively high overvoltage (8 V above the breakdown voltage $V_{BR}$ where signals begin).
\begin{figure}[ht]
    \centering
    \includegraphics[width=0.6\textwidth]{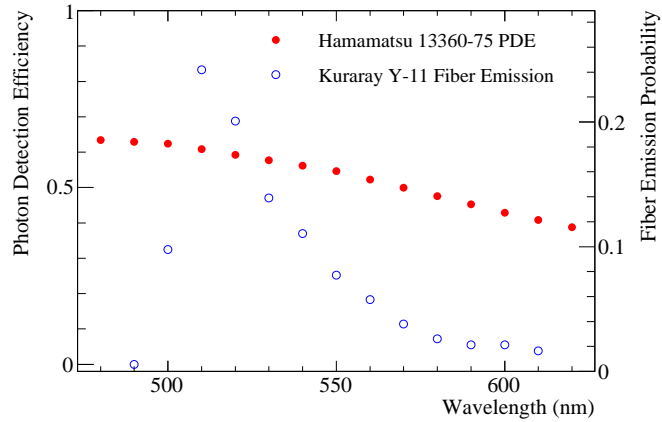}
    \caption{Photon detection efficiency vs wavelength ($\textrm{PDE}(\lambda)$) of a Hamamatsu 13360-75 SiPM at $V_{BR}+8 V$ and the emission spectrum for Kuraray Y-11 optical fibers at 1.5 m.}
    \label{fig:sipmpde}
    \end{figure} 
At this operating voltage, the PDE (figure \ref{fig:sipmpde}) is maximized ($\sim1.3$ times greater than standard operating voltage) and its dependence on voltage, temperature, and sensor variation is minimized \cite{Nepomuk}. Finally, this technique prevents gain dependence on voltage, temperature, and sensor variability from impacting the signal amplitude.
This technique is especially appropriate in the low-light conditions of the nEDM@SNS experiment. The probability that an individual sensor will see any photons in an event are small, so the PDE loss suffered when multiple photons from the same event hit the same SiPM is only a few percent.

\subsection{Simulation procedure}
\label{sec:simprocedure}
To quantify the expected light collection and event detection efficiency, as well as investigate the effects of some system parameters, GEANT4 was used to track photons from their initial position until they were wavelength-shifted in the fibers. We avoid simulating the transmission of fibers and detection in the SiPMs by applying ad hoc transmission and detection probabilities.

GEANT4 was a natural fit for the light collection simulation because of its thorough treatment of optical physics and material properties \cite{Gumplinger}.
The simulation is constructed as follows: 80 nm photons are emitted uniformly in the superfluid \textsuperscript{4}He of a measurement cell. They travel through the helium until reaching the dTPB layer on the measurement cell wall, which absorbs and re-emits 33\% of the photons wavelength-shifted in a spectrum centered around 450 nm \cite{Gehman2}.
Photons headed toward the cell wall instrumented with optical fibers pass through the acrylic with a small chance of being absorbed before reaching the fibers, where they are collected, wavelength-shifted (spectrum shown in figure \ref{fig:sipmpde}), and then killed by the simulation. A Vikuiti dielectric mirror (96\% reflectivity \cite{Mumm}) surface is installed on the non-fiber cell wall surfaces to prevent photons from escaping the measurement cell.
    \begin{table}[h!]
\begin{center}
\caption{Assumed fiber transport photon efficiency probabilities.}
\begin{tabular}{ |l|l| }
  \hline
  \multicolumn{2}{|c|}{Efficiencies in Fiber Transport} \\
  \hline
  Fiber trapping & $11\%$ \cite{Kuraray} \\
  Fiber bends & $95\%$ \cite{Kuraray} \\
  WLS Fiber transmission & $53\%$ \cite{Mu2e,Nova} \\
  Clear Fiber transmission & $74\%$ \cite{Pofeska} \\
  WLS/clear fiber interface & $90\%$ \cite{Kriplani} \\
  Fiber/SiPM interface & $90\%$ \cite{Kriplani} \\
  \hline
  Total & $3.32\%$ \\
  \hline
\end{tabular}
\label{table:fibtrans}
\end{center}
\end{table}

The probability of a photon to reach a SiPM once it has been successfully wavelength-shifted in a fiber is determined by the product of the Y-11 fiber trapping efficiency \cite{Kuraray}, transmission efficiencies in the Y-11 \cite{Mu2e, Nova} and clear fibers \cite{Pofeska}, and losses from fiber transitions and bends \cite{Kriplani} (See Table \ref{table:fibtrans}). A photon is considered detected if it reaches a SiPM and a uniform random number R within $0-1 < \textrm{PDE}(\lambda)$ (figure \ref{fig:sipmpde}). The photon counting technique outlined in Sec. \ref{sec:lightcollectionapp} is then accounted for by limiting each SiPM in the array to one photon per event.
\section{Results and Discussion} \label{sec:results}
\subsection{Parametric studies}
Before capture events were systematically studied in GEANT4, preliminary simulations were performed to study parametric effects from the optical fibers, dTPB, and Vikuiti VM2000 dielectric film. We use the fiber collection efficiency i.e., the fraction of photons that are successfully wavelength-shifted in the fibers, as the metric for these effects.

Vikuiti was defined using a native GEANT4 ``rough'' optical surface.  
\begin{figure}[h]
\centering
    \includegraphics[width=0.6\textwidth]{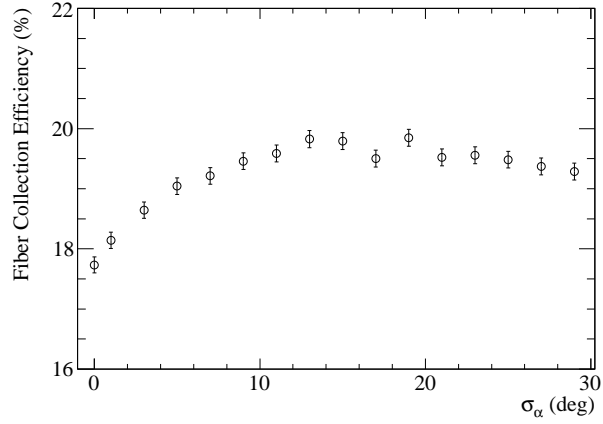}
    \caption{Fiber collection efficiency vs. Vikuiti specularity.}
    \label{fig:SigmaAlpha}
    \end{figure}
Photons incident on the optical surface undergo specular lobe reflection with a spread parameterized by $\sigma_\alpha$, which represents the width of a Gaussian distribution of surface micro-facet normals.
Fiber collection efficiency does not depend strongly on $\sigma_\alpha$ (figure \ref{fig:SigmaAlpha}). While the value of $\sigma_\alpha$ has not been experimentally measured, Vikuiti is known to not be a perfect specular reflector ($\sigma_\alpha > 0$). Due to the relative independence of light collection efficiency with $\sigma_\alpha$, it is taken to be $3 ^{\circ}$, the median value between a perfect specular reflector and maximum efficiency, with $<5 \%$ uncertainty.

\begin{figure}[h]
    \centering
    \includegraphics[width=0.6\textwidth]{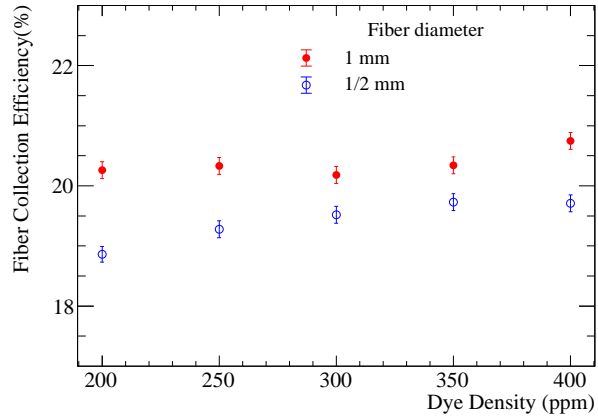}
    \caption{Fiber collection efficiency vs. dye density for different fiber diameters.}
    \label{fig:DyeDensity}
    \end{figure}
One method for increasing the fiber collection efficiency is to use higher concentration of the WLS dye in the fibers. By concentrating the dye, the probability of a photon shifting wavelength while traversing a fiber increases. This is tested in our simulation framework by scaling the wavelength-dependent absorption length of the fibers according to the dye density,
\begin{equation} \label{eq:3}
\lambda_{a} = \frac{1}{A_{r}(\lambda)\rho_{d} \ln{10}} ,
 \end{equation} 
where $A_{r}(\lambda)$ is the absorbance \cite{Kuraray} and $\rho_{d}$ is the dye density. The resulting simulations indicate that the fiber collection efficiency is only weakly dependent on the dye density in this light collection system, (figure \ref{fig:DyeDensity}) suggesting the fibers can be used at their nominal concentration of 200 ppm. 

The impact of fiber diameter on collection efficiency was also investigated by simulating fibers with \sfrac{1}{2} mm and 1 mm diameters, both of which are available from Kuraray. Due to mechanical tolerances, the SiPMs should be somewhat larger than the clear fibers, which should in turn be somewhat larger than the WLS fibers. Larger differences ease mechanical tolerance requirements, but since the dark rate is proportional to sensor area the sensors cannot be made arbitrarily large. The chosen SiPMs are 1.3 mm on a side. The next-largest sensor is 3 mm on a side; five times the area and five times the dark rate. There is only $\sim$ 5\% performance difference for the different fiber diameters (figure \ref{fig:DyeDensity}) so \sfrac{1}{2} mm diameter was chosen to relax mechanical tolerances.

Another idea for increasing the collection efficiency is to embed the fibers in the acrylic of the cell walls. Embedding the fibers reduces the change in the index of refraction at the fiber/cell wall interface by eliminating the helium gap. This should decrease the number of photons that are prevented from reaching the fibers because of total internal reflection off this boundary.

\begin{figure}[h]
    \centering
    \includegraphics[width=0.6\textwidth]{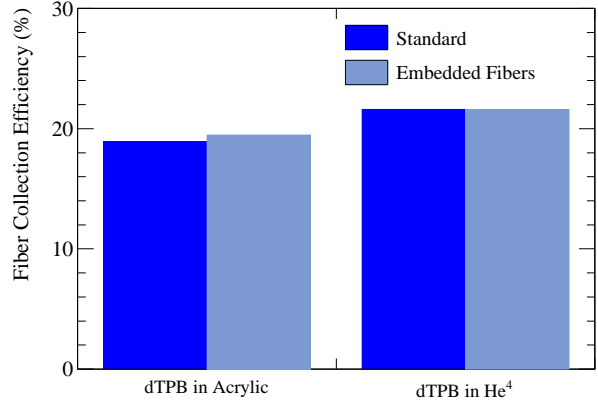}
    \caption{Fiber collection efficiency for embedded vs. unembedded fiber configurations. "dTPB in Acrylic" assumes that dTPB light is created inside the acrylic; "dTPB in He$^{4}$" assumes that dTPB light is created outside. }
    \label{fig:ParametricStudies}
    \end{figure}
Fiber embedding was tested in four scenarios: with dTPB light produced within the acrylic on the measurement cell wall boundary or outside of the acrylic (on the \textsuperscript{4}He side of the boundary) and with fibers embedded or unembedded. As seen in figure \ref{fig:ParametricStudies}, there is little dependence of fiber collection efficiency on these parameters. Due to the mechanical challenges of embedding the fibers, we adopt the unembedded geometry moving forward.

\subsection{UCN-\textsuperscript{3}He Capture Events}
\label{sec:capture}
Our optimized geometry consists of 192 \sfrac{1}{2} mm diameter unembedded fibers with 200 ppm dye density. Each fiber is looped into four adjacent segments along the cell wall with both fiber ends read out by individual SiPMs, leading to 384 total SiPMs. $\sigma_\alpha$ for the Vikuiti VM2000 reflector film was set to $3 ^{\circ}$.  
\begin{figure}[h]
\centering
    \includegraphics[width=0.6\textwidth]{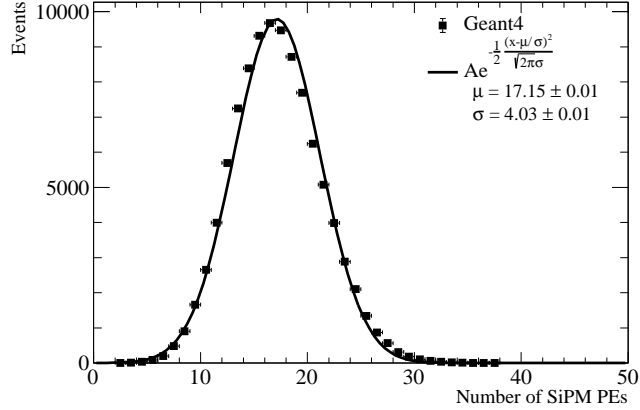}
    \caption{Number of photoelectrons detected in SiPMs from UCN-\textsuperscript{3}He capture events distributed uniformly throughout the measurement cell volume.}
    \label{fig:lightcollection1}
    \end{figure}
    
Ito et al. \cite{Ito2} predict approximately 5050 primary UV photons from capture events when the electric field is 40 kV/cm. 
Taking the product of the efficiencies in Table \ref{table:fibtrans}, $3.32\%$ of green photons produced in the WLS fibers will be transmitted to the SiPM readout.
Thus, for this simulation which tracks only up until green photons are produced in the fibers, UCN-\textsuperscript{3}He capture events are simulated by emitting 167 $\pm$ 12.9 EUV photons from a capture event position into 4$\pi$. 
These photons then undergo the process described in section \ref{sec:simprocedure}. A simulation of 100,000 capture events  emitted uniformly throughout the cell found the mean number of detected photoelectrons (NPE) to be
$17.15 \pm 4.03$ per event (figure \ref{fig:lightcollection1}).
    
\subsection{Event location reconstruction}
\label{sec:reconstruction}
\begin{figure}[h]
\centering
    \includegraphics[width=0.6\textwidth]{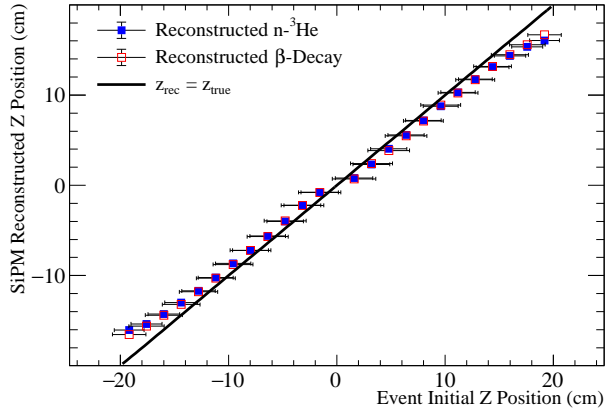}
    \caption{Reconstructed vs. true position along the beam axis ($z_{rec}$ vs. $z_{true}$) of UCN-\textsuperscript{3}He (blue) and beta decay (red) events using a truncated weighted mean of hit readout channel positions.}
    \label{fig:eventlocationreconstruction}
    \end{figure}
Each readout channel (a group of 16 SiPMs) collects photons from fibers distributed over a strip of the cell wall limited to a 3.2 cm range in the beam direction. 
Thus, the distribution of hits binned by readout channel provides information on the event position in the beam direction. This information can be utilized to identify systematic effects and, as seen in Sec (\ref{sec:cuts}), improve signal cuts that suppress experimental backgrounds. 
    
Event z-position is reconstructed by calculating the mean position of struck readout channels weighted by the NPE detected in each channel.
 At event positions near the cell windows ($|z| \gtrsim 15 $ cm), the reconstructed position is skewed by the asymmetric tail of the SiPM readout distribution towards the center of the cell. This effect is reduced by implementing a truncation method, in which readout channels with 1 photoelectron are removed before the weighted mean is calculated (figure \ref{fig:eventlocationreconstruction}).

\subsection{Background Suppression Cuts}
\label{sec:cuts}
The primary source of background events is beta decay of the stored UCN, which produces an optical signal comparable to that of the UCN-\textsuperscript{3}He capture events. Background contributions from neutron activation, UCN wall loss, and cosmic rays are smaller and not studied here.
\begin{figure}[htbp!]
\centering
    \includegraphics[width=0.6\textwidth]{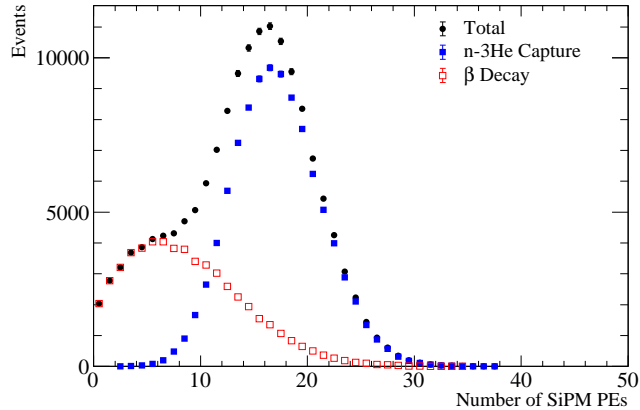}
    \caption{Photoelectron distribution of UCN-\textsuperscript{3}He capture (signal) and beta decay (background) events. The details of the beta decay simulation, including the relative number of beta/capture events, are described in the text.}
    \label{fig:lightcollectionbackgrounds}
    \end{figure}
    
Betas are selected from the neutron beta decay spectrum and simulated in the nEDM@SNS GEANT4 framework using a native scintillation process parameterized by the properties of liquid helium in a uniform electric field \cite{McKinsey2, Phan}. 
The distribution of NPEs detected per event from neutron beta decays is then normalized by the ratio of the beta decay rate to the capture rate at nominal wall loss, polarization loss, and \textsuperscript{3}He density, $\sfrac{R_{\beta}}{R_3} = 0.55$ \cite{Leung2} to produce figure \ref{fig:lightcollectionbackgrounds}. 
\begin{figure*}[htbp!]
    \centering
    \subfigure[]{\label{fig:NPEzDependence}\includegraphics[width=0.49\textwidth]{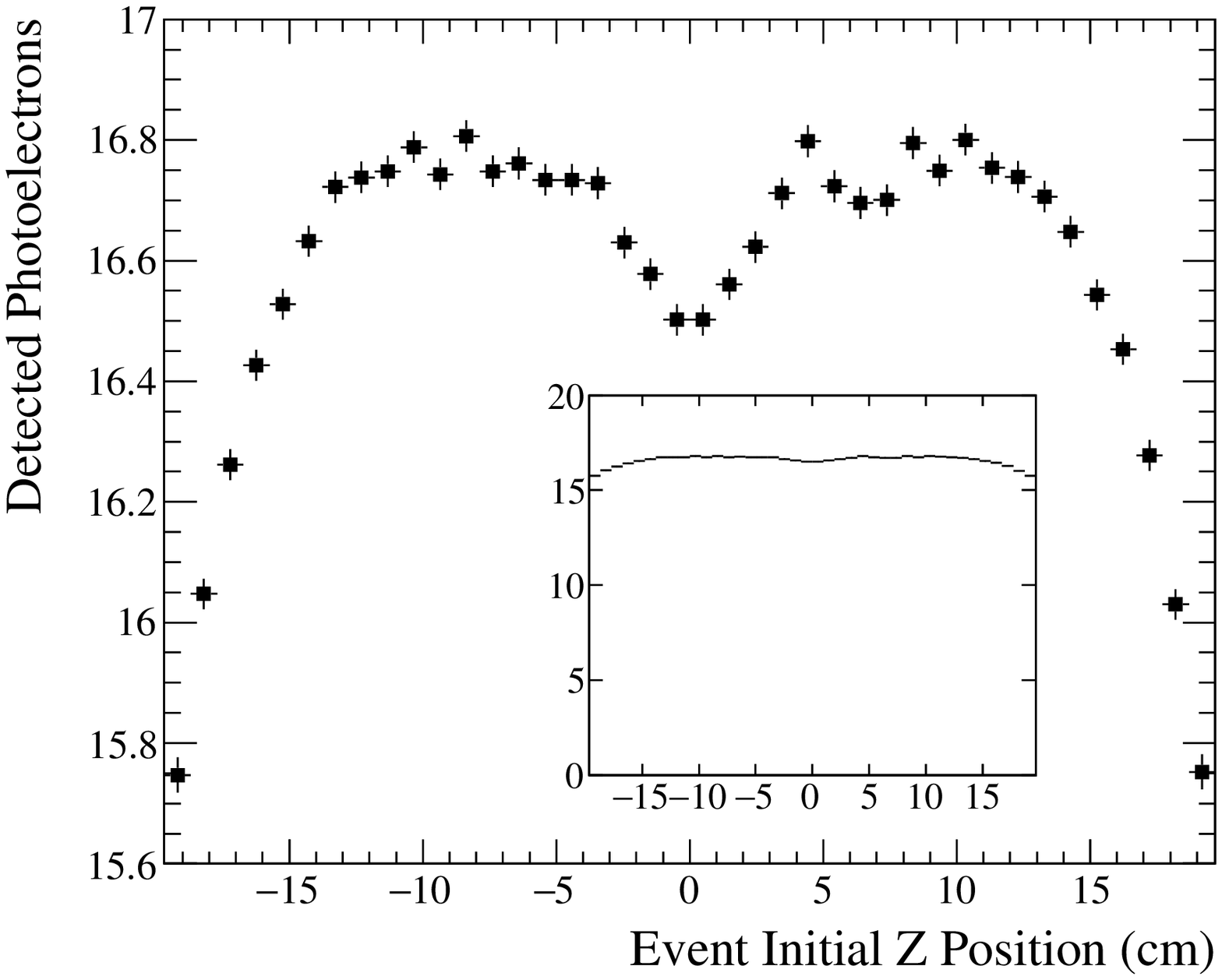}}
    \subfigure[]{\label{fig:NPExDependence}\includegraphics[width=0.49\textwidth]{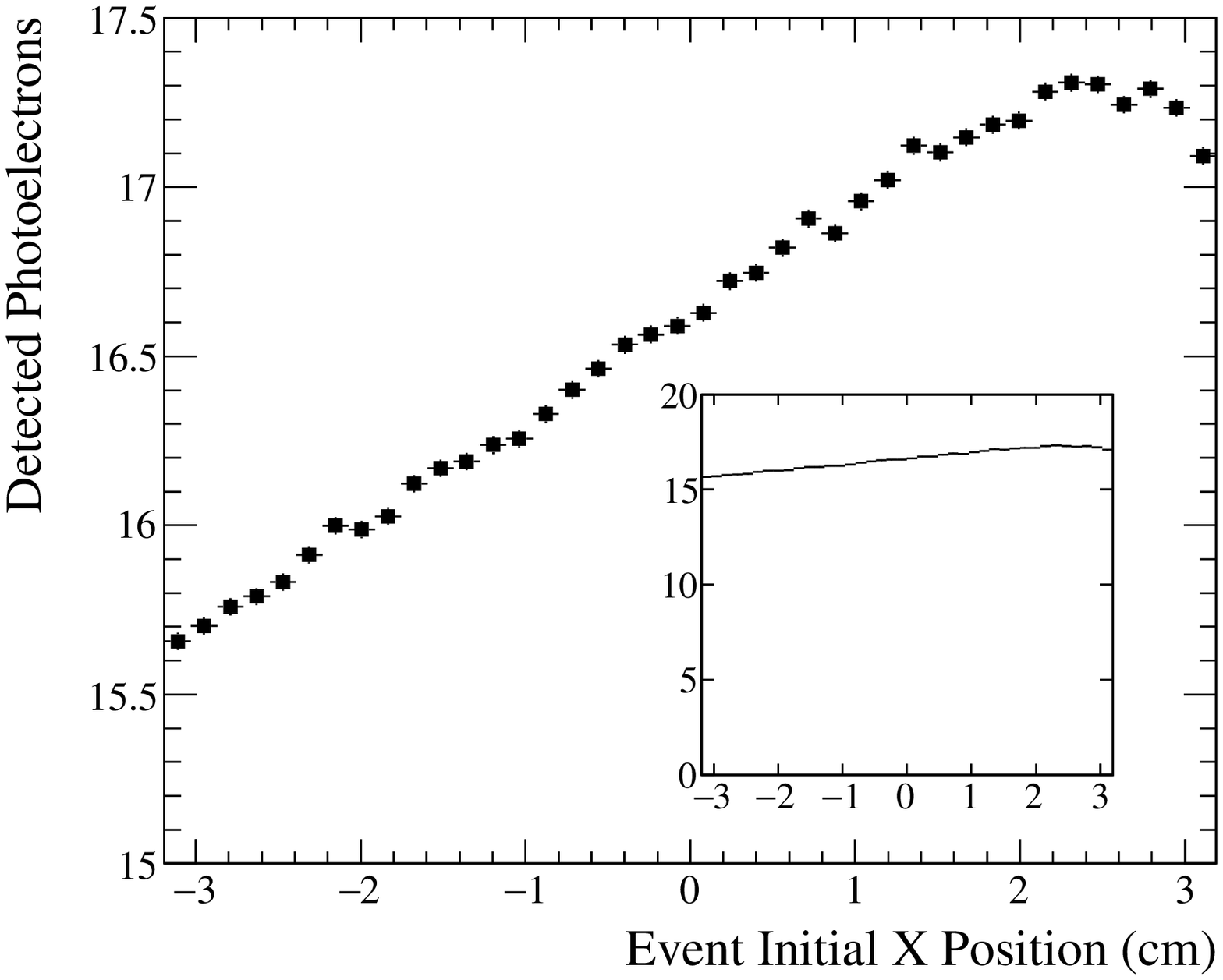}}
    \caption{$\langle \textrm{NPE} \rangle$ detected from UCN-\textsuperscript{3}He events (a) in the beam (z) direction, and (b) in the direction perpendicular to the beam and the fiber plane (x; fibers at x = +3.5 cm).}
    \label{fig:PositionDependencies}
\end{figure*}

In order to maximize signal sensitivity, the beta decay background (and other background sources with broad energy deposition distributions) can be suppressed via cuts on the NPE distribution relative to the expected yield from monoenergetic UCN-\textsuperscript{3}He capture events.
The design goal of the nEDM@SNS experiment is to reduce the beta decay background by a factor of two ($\zeta_{\beta} = \sfrac{1}{\epsilon_{\beta}} > 2$) while maintaining a high capture event detection efficiency ($\epsilon_{3} \geq 0.93$). 
This can be accomplished with relatively narrow NPE cuts about the capture peak, with an especially tight cut placed on the lower side of the capture peak where the beta signal is concentrated. The tight and asymmetric nature of the cuts about the capture peak, while maximizing rejection of the beta background,
introduce position-dependent event detection efficiencies since the signal amplitude is modestly position-dependent along both the beam direction and the x-direction (figure \ref{fig:PositionDependencies}).
\begin{figure*}[htbp!]
    \centering
    \subfigure[]{\label{fig:n3HeCaptureEfficiency}\includegraphics[width=0.49\textwidth]{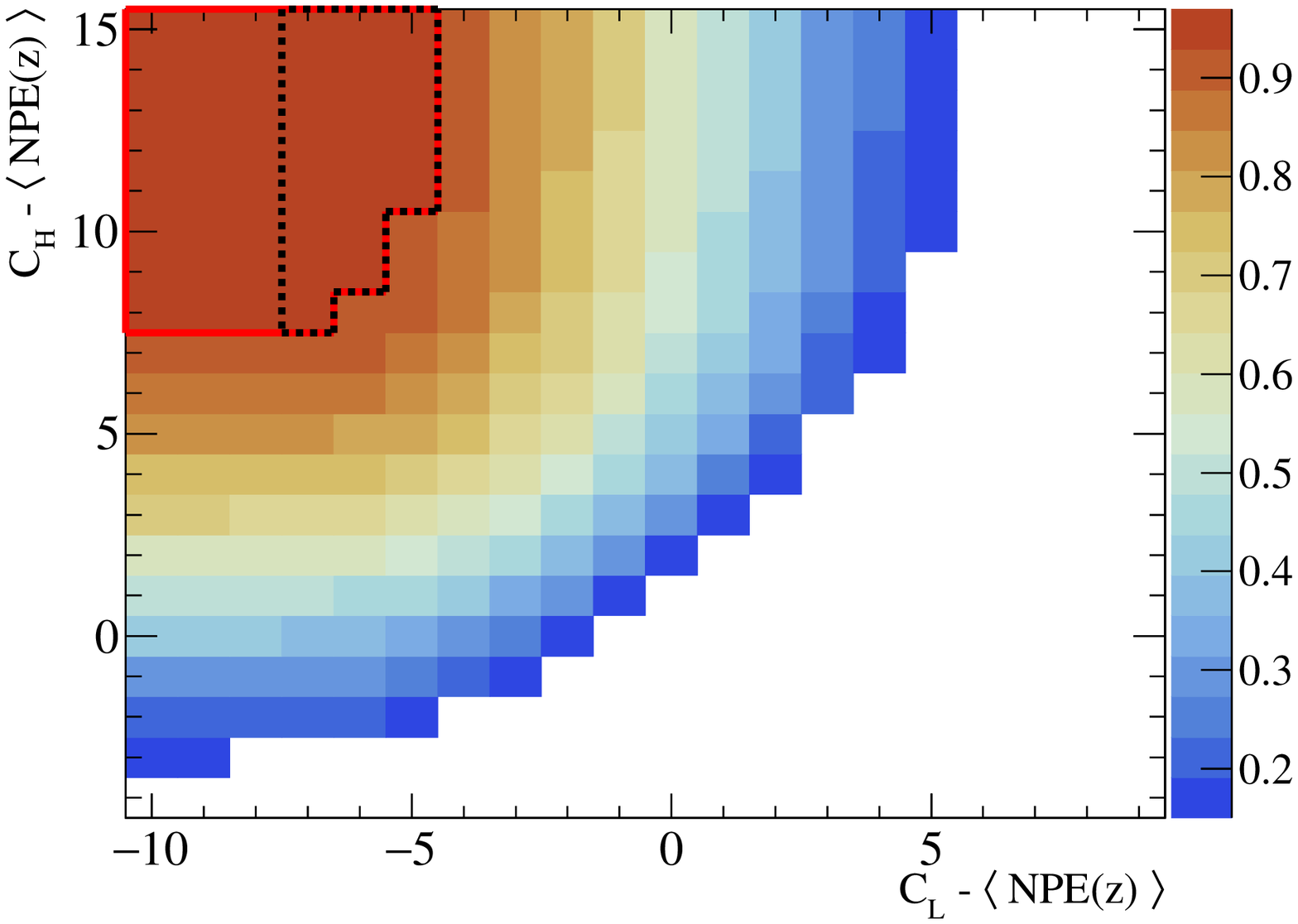}}
    \subfigure[]{\label{fig:BetaEfficiency}\includegraphics[width=0.49\textwidth]{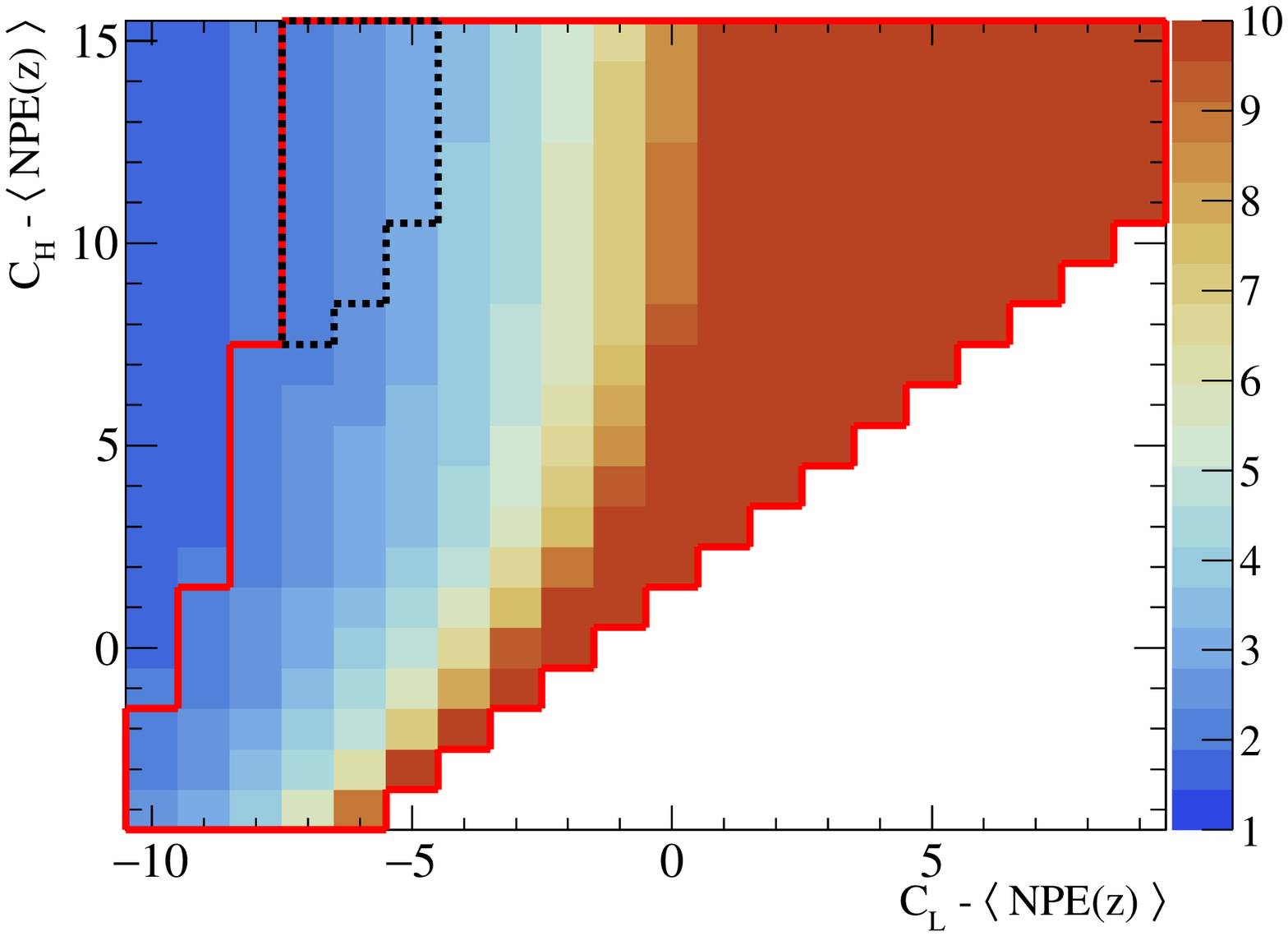}}
    \caption{Effect of NPE cut values on (a) UCN-\textsuperscript{3}He capture signal detection efficiency (b) beta decay background rejection factor. The region within the red contour in (a) satisfies $\epsilon_{3} \geq 0.93$ and in (b) satisfies $\zeta_{\beta} > 2$. The region within the dashed black contour satisfies both conditions.}
    \label{fig:CutOptimization}
\end{figure*}

The capture signal efficiency variation along the beam axis can be reduced to less than 1\% by using the reconstructed beam axis position (section \ref{sec:reconstruction}) to define z-dependent NPE cuts.
Figure \ref{fig:CutOptimization} shows $\epsilon_{3}$ (left) and $\zeta_{\beta}$ (right) as a function of the NPE cut lower bound ($C_L$) and upper bound ($C_H$) defined relative to the average number of photons for the  event z-position ($\langle \textrm{NPE(z)} \rangle$). The regions outlined in red show the cut combinations that satisfy at least one of the conditions $\epsilon_{3} \geq 0.93$ or $\zeta_{\beta} > 2$. The region outlined in the dashed black line defines the set of cut combinations that satisfies both conditions.

\begin{figure*}[htbp!]
    \centering
    \subfigure[]{\label{fig:cut}\includegraphics[width=0.49\textwidth]{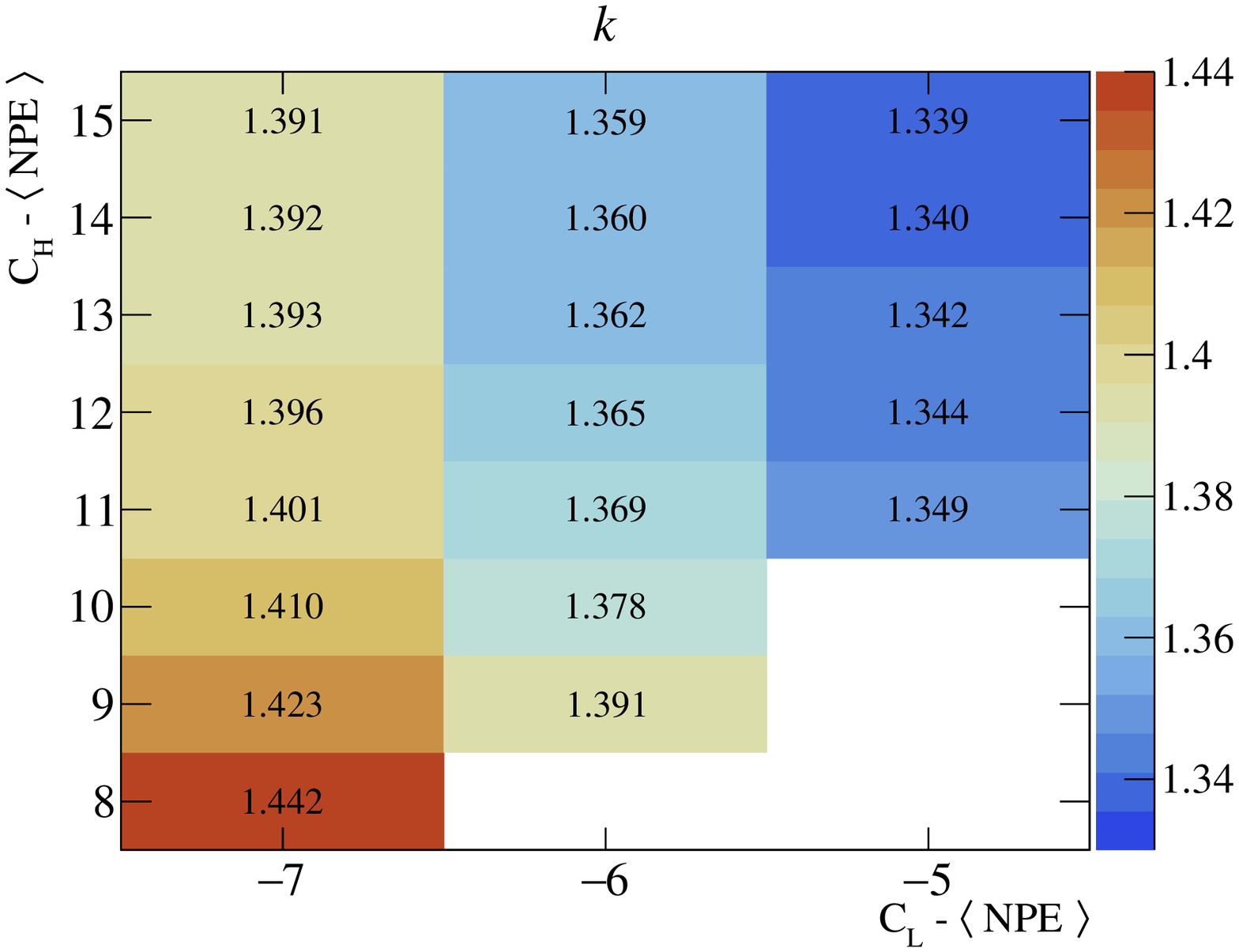}}
    \subfigure[]{\label{fig:cut1}\includegraphics[width=0.49\textwidth]{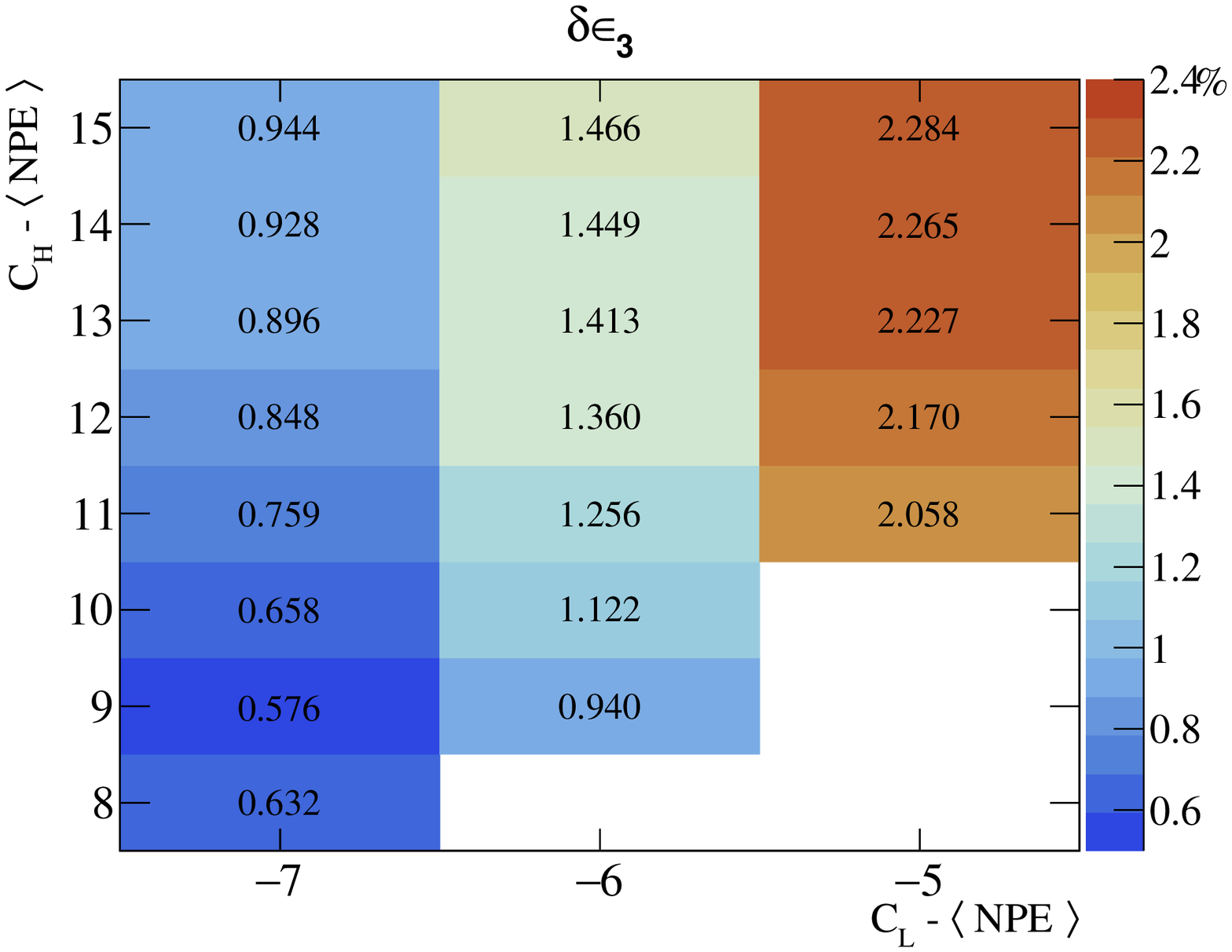}}
    \subfigure[]{\label{fig:cut2}\includegraphics[width=0.49\textwidth]{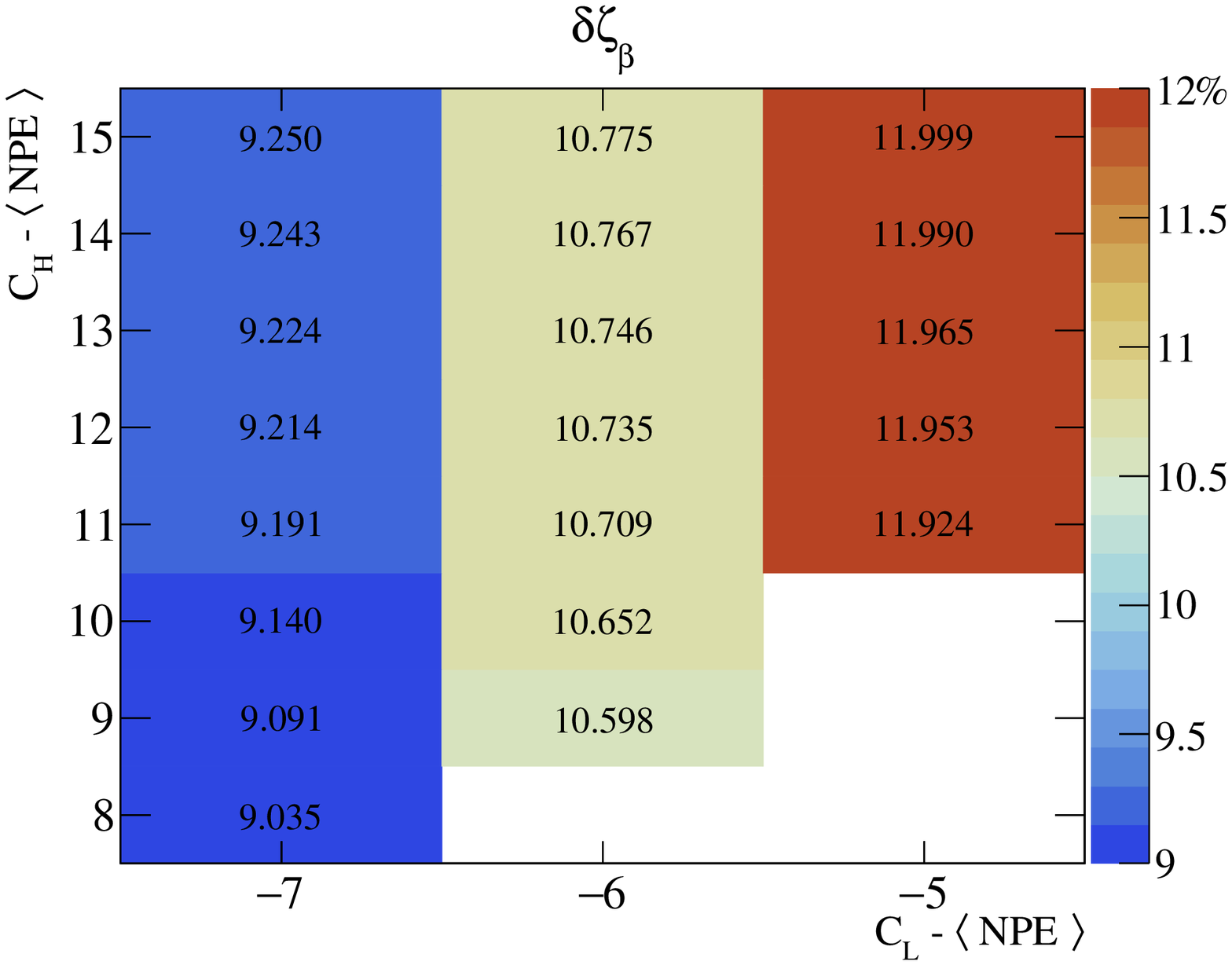}}
    \caption{NPE cut placement optimization parameters in region where $\epsilon_{3} \geq 0.93$, $\zeta_{\beta} > 2$. The color scale is (a) k (see eq. \ref{eq:4}) (b) the signal detection efficiency variation across the measurement cell ($\delta \epsilon_{3}$) (c) the beta decay background rejection variation across the measurement cell ($\delta \zeta_{\beta}$).}
    \label{fig:cutfluctuations}
\end{figure*}

\begin{figure*}[htbp!]
    \centering
    \subfigure[]{\label{fig:n3HeCaptureEfficiency}\includegraphics[width=0.49\textwidth]{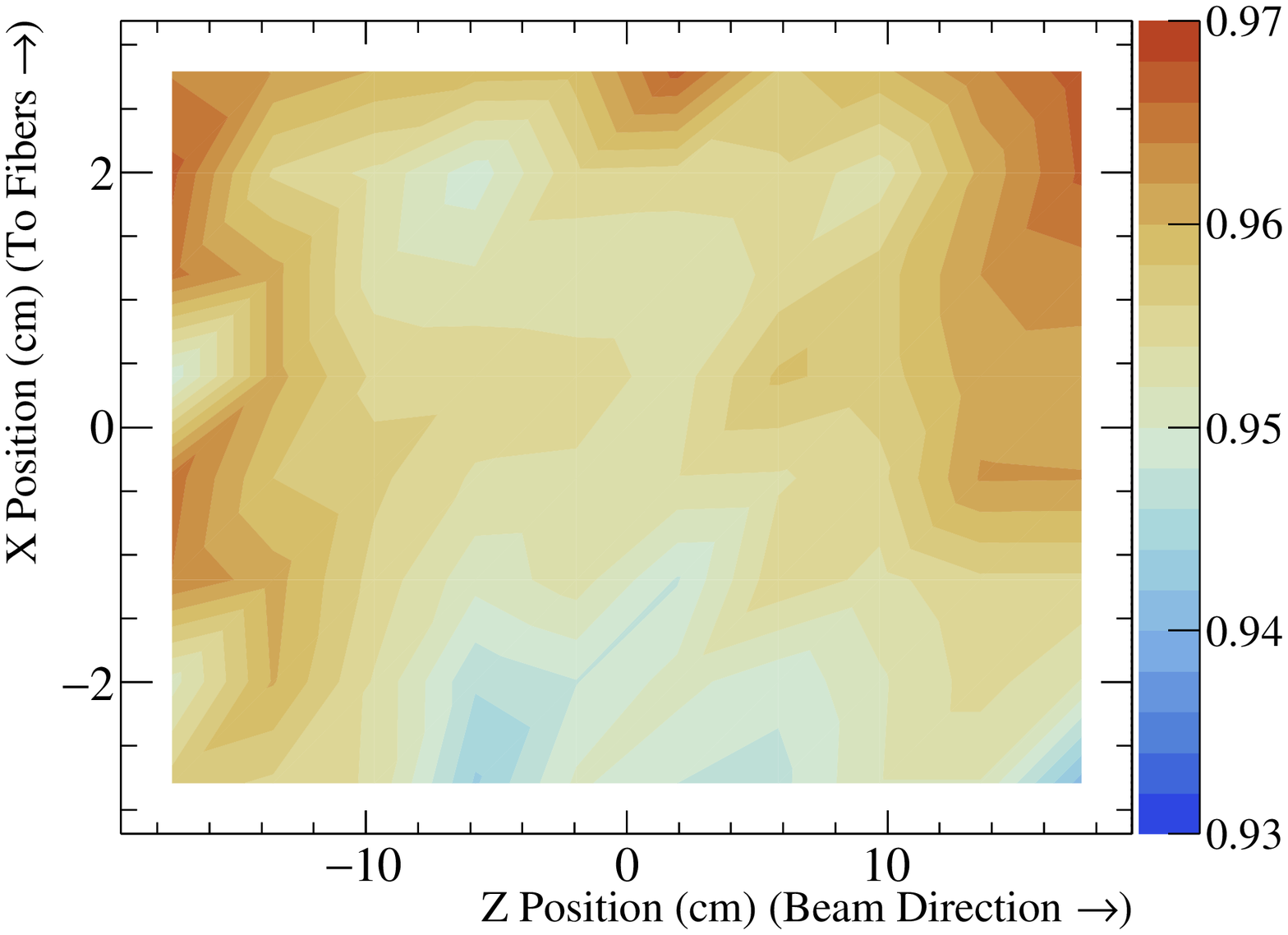}}
    \subfigure[]{\label{fig:BetaEfficiency}\includegraphics[width=0.49\textwidth]{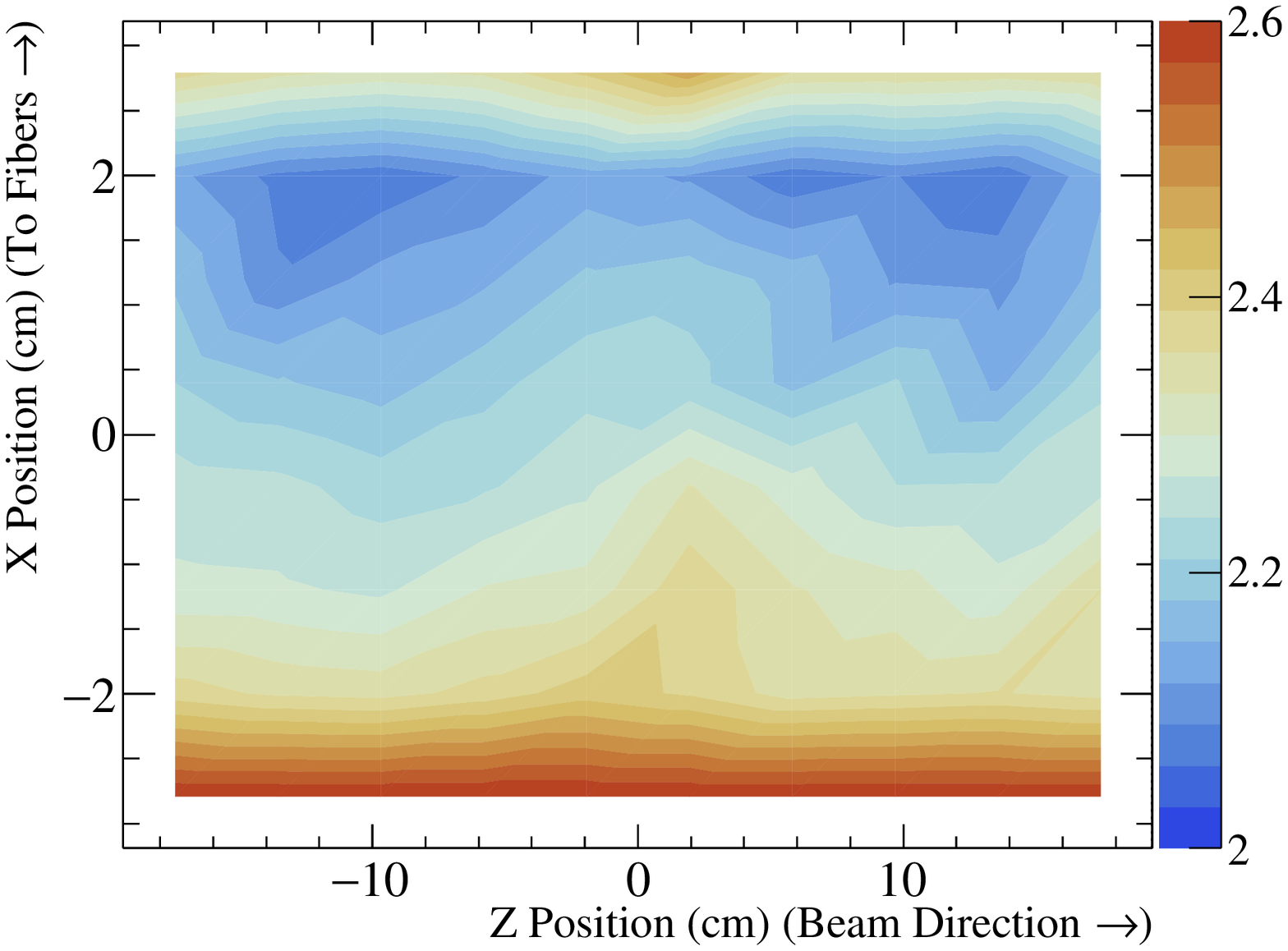}}
    \caption{(a) Capture event detection efficiency and (b) beta decay background rejection factor with optimal $\langle \textrm{NPE(z)} \rangle$ cuts throughout the measurement cell.}
    \label{fig:Efficiencies}
\end{figure*}

No x-position information is available, so the analogous efficiency variation in that coordinate cannot be addressed with position-dependent cuts. Instead, simulations were used to select optimal cuts from the 20 cut combinations that satisfy the $\epsilon_{3} \geq 0.93$ and $\zeta_{\beta} > 2 $ requirements. We simulated each cut combination independently, calculating the variation in signal detection efficiency ($\delta \epsilon_{3}$) and background rejection ($\delta \zeta_{\beta}$) as the normalized root mean square deviation
across the measurement cell. Additionally, we calculated the fluctuation in the detected capture signal ($\sigma_{S_{tot}} \textrm{/} S_{tot} = k \sqrt{1/S_{tot}}$) where
\begin{equation} \label{eq:4}
k = \sqrt{\frac{1}{\epsilon_3}} \sqrt{1+2\frac{\epsilon_{\beta} B_{tot}}{\epsilon_3 S_{tot}}},
 \end{equation} quantifies the degradation of the signal count uncertainty relative to the background-free case for different background suppression cuts. $S_{tot}$ and $B_{tot}$ are the total signal and background NPE counts, respectively. Figure \ref{fig:cutfluctuations} shows the results for the specified 20 cut combinations. $C_L$ = $\langle \textrm{NPE(z)} \rangle$ – 7 and $C_H$ = $\langle \textrm{NPE(z)} \rangle$ + 9 are selected as the best trade-off between minimizing $k$, $\delta \epsilon_{3}$, and $\delta \zeta_{\beta}$. Figure \ref{fig:Efficiencies} shows $\epsilon_3$ (left) and $\zeta_{\beta}$ (right) vs. cell position with these cuts. Over the entire volume of the cell, $\epsilon_{3}$ and $\zeta_{\beta}$ meet their design goals with variations $\delta \epsilon_{3} < 1\%$, and $\delta \zeta_{\beta}$ < 10\%.

The $\zeta_\beta$ x-dependence is the combined effect of the photoelectron detection efficiency x-dependence (figure  \ref{fig:NPExDependence}), the asymmetry of the NPE distribution for beta events (figure \ref{fig:lightcollectionbackgrounds}) and the relatively long pathlength ($\mathcal{O}$(2 cm)) of beta particles in liquid \textsuperscript{4}He. The result of this last effect is that betas produced near the cell walls may only deposit a fraction of their energy in the \textsuperscript{4}He, reducing their scintillation signal in a way that can depend on the direction of the electric field. This could conceivably result in different signal to background ratios for different electric field directions, which could indirectly lead to small shifts in the extracted frequency parameter for different electric field directions, and thus, a false EDM. This effect was investigated by repeating the previous simulations after reversing the electric field. 
\begin{figure}[h]
\centering
    \includegraphics[width=0.6\textwidth]{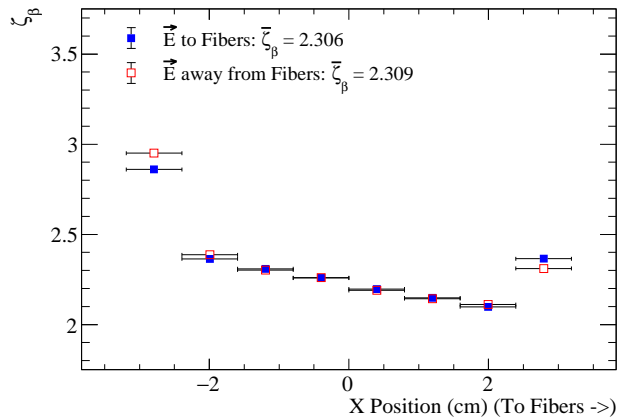}
    \caption{Beta decay background rejection vs. x-position for the two electric field configurations.}
    \label{fig:efielddep}
    \end{figure}
Figure \ref{fig:efielddep} shows a small (<1\%) E-field direction dependence of $\zeta_\beta$ integrated over the cell's volume. With no mitigating effects this would lead to a false nEDM of $\sim 1 \times 10^{-28}$ e cm \cite{Leung2}. This is already acceptable, but any frequency shift can be easily suppressed by allowing the signal-to-background ratio to vary in the frequency extraction fit rather than assuming a fixed value. Further cancellation is provided by the nEDM@SNS two-cell design: a true EDM will result in opposite sign frequency shifts for the two cells since they have opposite values of $\vec{E} \cdot \vec{B}$. In contrast, any frequency shift resulting from a field-dependent signal-to-background ratio depends on $\vec{E} \cdot \hat{f}$ ($\hat{f}$ is the normal vector of the fiber plane relative to the cell interior) which is the same for both cells.
    
This small false nEDM, together with the simulated detection efficiencies, sufficient $\langle \textrm{NPE} \rangle$ and negligible dependencies on the parameters addressed in this study, suggest that the current design of the light collection system will meet the requirements of the experiment. Uncertainties from ad hoc inputs into the simulation are addressed by using known worst case values, suggesting that the experimental performance of the light collection system could exceed the results found by this simulation study. Additionally, an increase in the PDE, and thus the $\langle \textrm{NPE} \rangle$, could be foreseen with improvements in SiPM devices in the forthcoming years. Although their contribution to the total background is less significant than beta decay, a worthwhile extension to this study would properly model the effect of neutron activation and UCN wall loss on the signal detection efficiency and background rejection capability of the system.

\section{Summary} \label{sec:conclusions}
The light collection system responsible for reading out the signal from UCN-\textsuperscript{3}He capture scintillation in the nEDM@SNS experiment has been simulated. This system consists of a WLS dTPB coating, a thin film dielectric reflector, WLS optical fibers, and silicon photomultipliers.

Simulations indicate that previously untested parameters, such as fiber diameter, fiber dye density, fiber embedding, and thin film reflector surface roughness, have only minor effects on the final results. A mean of $\sim17$ photoelectrons are detected per UCN-\textsuperscript{3}He capture event and the position of the event along the beam direction can be reconstructed to within several cm. Position-dependent cuts on the number of detected photoelectrons reduce the beta decay background to sufficiently low levels. Additionally, the event position information can be used to identify further backgrounds/systematic effects. With optimized cuts, the efficiency of capture event detection is uniform to < 1\% across the measurement cell and the efficiency of beta decay background rejection is uniform to < 10\%. These results hold promise for the successful implementation of the light collection system in the upcoming nEDM@SNS experiment.

\acknowledgments

The authors would like to thank John Ramsey for the set of light collection system CAD models and Takeyasu Ito and Kent Leung for their support in the development of the light collection system simulation. The authors would also like  to thank Leah Broussard, Bob Golub, Steve Lamoreaux, and Mike Hayden for their discussion and suggestions. This work was supported in part by the U.S. Department of Energy, Office of Science, Office of Workforce Development for Teachers and Scientists (WDTS) under the Science Undergraduate Laboratory Internship program.

\end{document}